\documentclass[aps,twocolumn,prd,preprintnumbers,amsmath,amssymb,superscriptaddress,nofootinbib]{revtex4-2}
\pdfoutput=1

\usepackage[export]{adjustbox}

 \usepackage{color}
 \usepackage{epsfig}
 \usepackage{ulem}

\usepackage[pdftex,bookmarks,linktocpage,pdfpagelabels,plainpages=false,hyperfigures,linkcolor=blue,citecolor=blue]{hyperref}
\usepackage{cleveref}
\hypersetup{colorlinks=true}

\definecolor{White}{rgb}{1,1,1}
\definecolor{Red}{rgb}{1,0.1,0}
\definecolor{LightYellow}{rgb}{1,1,.875}
\definecolor{SteelBlue}{rgb}{.273,.508,.703}
\definecolor{navy}{rgb}{0,0,.5}
\definecolor{LightCyan}{rgb}{.875,1,1}
\definecolor{DarkRed}{rgb}{.543,0,0}
\definecolor{HotPink}{rgb}{1,.41,.70}
\definecolor{ForestGreen}{rgb}{.13,.54,.13}
\definecolor{OliveDrab}{rgb}{.42,.55,.14}
\definecolor{MediumBlue}{rgb}{0,0,.80}
\definecolor{RoyalBlue}{rgb}{.25,.41,.88}
\definecolor{DeepSkyBlue}{rgb}{0,.746,1}
\definecolor{Brown}{rgb}{0.545,0.271,0.074}
\definecolor{Purple}{rgb}{0.637,0.285,0.641}

\newcommand{\dis}[1]{\begin{equation}\begin{split}#1\end{split}\end{equation}}
\def\bea{\begin{eqnarray}}
\def\eea{\end{eqnarray}}

\begin{document}

\preprint{CTPU-PTC-22-12} 
 
\title{Axion dark matter with thermal friction}

\author{Kiwoon Choi}
\email{kchoi@ibs.re.kr}
\author{Sang Hui Im}
\email{imsanghui@ibs.re.kr}
\author{Hee Jung Kim}
\email{heejungkim@ibs.re.kr}
\affiliation{Center for Theoretical Physics of the Universe,   Institute for Basic Science, Daejeon 34126,   Korea }
\author{Hyeonseok Seong}
\email{hyeonseok.seong@desy.de}
\affiliation{Center for Theoretical Physics of the Universe,   Institute for Basic Science, Daejeon 34126,   Korea }
\affiliation{Deutsches Elektronen-Synchrotron DESY, Notkestr. 85, 22607 Hamburg, Germany}

\begin{abstract}

Cosmological evolution of axion field in the early universe might be significantly affected by  a thermal friction induced by the axion coupling to thermalized hidden sector. We examine the effects of such a thermal friction on axion dark matter density and its perturbation when the thermal friction dominates over the Hubble friction until when the axion field begins to oscillate around the potential minimum.
We show that in the presence of sizable thermal friction there can be an exponential decay phase of the axion field before the oscillation phase,  during which the axion energy density is efficiently dissipated into hidden thermal bath.
Consequently, the previously excluded parameter region due to overclosing relic axion density becomes cosmologically viable with thermal friction.
In particular, a QCD axion much lighter than $\mu$eV is viable without tuning the initial misalignment angle. We also find that thermal friction can affect the density perturbation of axion dark matter in various ways.
For instance, it can alleviate the large-scale isocurvature bound on axion dark matter in the pre-inflationary PQ breaking scenario, which would 
make the pre-inflationary axion dark matter compatible with high scale inflation over a wide range of model parameters. 
In the post-inflationary PQ breaking scenario, thermal friction
can also significantly change the scaling behavior of axionic strings, and therefore the typical size of the resultant axion miniclusters.

\end{abstract}  

\maketitle

\section{Introduction}

The QCD axion is among the most compelling candidates for physics beyond the standard model (BSM). It solves 
the strong CP problem \cite{Peccei:1977hh, Weinberg:1977ma, Wilczek:1977pj} and provides a natural candidate for cold dark matter \cite{Preskill:1982cy, Abbott:1982af, Dine:1982ah}. 
Similar particles, dubbed axion-like particles (ALPs), are ubiquitous in string theory \cite{Svrcek:2006yi, Arvanitaki:2009fg}, some of which can be identified as ultralight dark matter with a broad mass range and interesting cosmological consequences \cite{Arias:2012az, Grin:2019mub,Choi:2020rgn}.   
Usually the abundance of axion dark matter is determined by the axion mass and decay constant depending upon the cosmological epoch when the Peccei-Quinn (PQ) symmetry breaking takes place \cite{Arias:2012az, Kawasaki:2014sqa}.
 
Axions can have a sizable interaction with a thermal system in the early universe. Such an interaction can make freeze-in production of axions if the interaction rate is weaker than the Hubble expansion rate, while it thermalizes axions in the other case that the interaction rate is greater than the Hubble expansion rate \cite{Graf:2010tv, Salvio:2013iaa, Bae:2011jb, Im:2019iwd}. The interaction with a thermal system can also give rise to an intriguing influence to the dynamics of the classical axion field through the form of macroscopic friction \cite{McLerran:1990de}. Possible effects of thermal friction on the dynamics of axion field  have been studied before in the context of warm inflation \cite{Mishra:2011vh, Visinelli:2011jy, Kamali:2019ppi, Berghaus:2019whh}, axionic dark energy \cite{Berghaus:2020ekh}, axionic early dark energy which may ameliorate the $H_0$ tension \cite{Berghaus:2019cls}, and electroweak baryogenesis by axion dark matter \cite{Im:2021xoy}.

In this work we explore the influence of thermal friction on the abundance of axion dark matter and also on the axion density perturbation. We will show that in the strong friction limit there is an exponential decay phase in the cosmological axion field evolution before the axion starts to oscillate around the potential minimum. During this phase the axion energy density is efficiently dissipated into a thermal bath, and the resultant axion dark matter density is significantly reduced compared with the conventional cosmological scenarios without thermal friction.\footnote{See e.g. \cite{Allali:2022yvx} for an alternative scenario where the axion abundance is significantly modified due to interaction with a light hidden sector.}
 This decay phase has been partially studied in different contexts including early dark energy \cite{Berghaus:2019cls} and electroweak baryogenesis by axion dark matter \cite{Im:2021xoy}.
Here we provide approximate analytic solutions for the axion field evolution in the presence of significant thermal friction 
in order to study its impact on axion dark matter density. 
The thermal friction-induced exponential decay of the axion field  allows for instance 
a QCD axion much lighter than $\mu$eV to be cosmologically viable without tuning the initial misalignment angle in the pre-inflationary PQ breaking scenario, and also a QCD axion much lighter than $0.1$\,meV to be viable in the post-inflationary PQ breaking scenario.

We also find that thermal friction can significantly affect the density perturbation of axion dark matter in various ways. For instance, it can alleviate the large-scale isocurvature bound on axion dark matter in the pre-inflationary PQ breaking scenario.
It also suppresses the power spectrum of axion density perturbation at small scales below a certain scale determined by the axion mass when the axion field underwent an exponential decay.  Moreover it turns off various resonance effects which can be present in the absence of thermal friction.

In the post-inflationary PQ breaking scenario, on the other hand, the thermal friction keeps the random axion field values in a scale even smaller than the horizon size. Thus it changes the scaling behavior of the axionic strings, making the strings to be present by $O(1)$ population in a scale significantly smaller than the horizon size. 
As a consequence, the typical axion minicluster size is to be smaller than the conventional one for a given axion mass. For the QCD axion, combined with the fact that the thermal friction allows a larger decay constant well above $10^{10}$ GeV in the post-inflationary PQ breaking scenario, the axion minicluster size can be much greater than the expected size in the conventional axion cosmology of negligible thermal friction.

The rest of this paper is organized as follows. 
In section \ref{sec:dynamics}, we discuss cosmological axion field dynamics in the presence of strong thermal friction in a general context and examine its impact on axion dark matter density. In section \ref{sec:qcdaxion}, we pay attention to the implications on the QCD axion dark matter. In section \ref{sec:hYM}, we discuss constraints and cosmological implications arising from the involved hidden thermal sector. In section \ref{sec:denp}, we study thermal friction effect on the density perturbation of axion dark matter. Section \ref{sec:conc} is for summary and conclusions.

\section{Cosmological axion evolution with thermal friction} \label{sec:dynamics}

Generic axion field may have a sizable interaction with hidden thermal system realized in the early universe. 
As a concrete example for thermalized hidden sector, 
we will consider $SU(N)$ Yang-Mills (YM) gauge fields without charged fermions.\footnote{ In the presence of  $SU(N)$-charged fermion lighter than about $\alpha_h^2 T_h$, 
the sphaleron transitions generating thermal friction for the axion field are suppressed due to 
the energy cost for producing light fermions, where $T_h$ is the temperature of the thermalized hidden YM sector
 \cite{McLerran:1990de, Berghaus:2020ekh, Im:2021xoy}. Thus we will assume that there is no light 
 $SU(N)$-charged fermion.
This also means that the QCD gluon fields would not be able to make a significant thermal effect on 
the evolution of axion field because of the light quarks.}
(For other examples, see e.g.~\cite{Mishra:2011vh}.) 
The relevant part of lagrangian for the axion field $\phi$ is given by
\dis{
{\cal L} \supset -\frac{1}{2} \partial^\mu \phi \partial_\mu \phi - V(\phi) +\frac{\alpha_h}{8\pi} \frac{\phi}{f_h} G^{\mu \nu a}_h \widetilde{G}_{h\, \mu \nu}^a \,, \label{lag}
}
where $G^a_{h\mu\nu}$ are the hidden YM field strengths, $\tilde G^a_{\mu\nu}=\frac{1}{2}\epsilon_{\mu\nu\rho\sigma}G^{a\rho\sigma}_h$ are their duals,  $\alpha_h=\frac{g_h^2}{4\pi}$ is the hidden YM fine structure constant, and the scale $f_h$ characterizes the interaction strength of the axion to the hidden YM fields.
For homogeneous and isotropic background, the axion field obeys the equation of motion:
 \dis{ 
\ddot{\phi} + 3H \dot{\phi} + V'(\phi)= -\frac{\alpha_h}{8\pi f_h} \langle G^{\mu \nu a}_h \widetilde{G}_{h\, \mu \nu}^a \rangle, \label{eom}
}
where the dot denotes time-derivative, the prime is the partial derivative with respect to $\phi$, $H$ is the Hubble expansion rate, and $\langle \cdots \rangle$ is the thermal average of the quantity. 
The thermal average of the hidden YM operator $G\tilde G$ gives rise to a friction term for the axion field dynamics by sphaleron processes \cite{McLerran:1990de, Berghaus:2020ekh}: 
\dis{
\frac{\alpha_h}{8\pi f_h} \langle G^{\mu \nu a}_h \widetilde{G}_{h\, \mu \nu}^a \rangle = \gamma_\phi \dot{\phi} + ...\label{eq:frictionterm}
}
with 
\dis{
\gamma_\phi \sim (N \alpha_h)^5 \frac{T_h^3}{f_h^2}.
}
Here $T_h$ is the temperature of the thermalized hidden sector, and the ellipsis stands for other terms 
which can be safely ignored in our discussion\footnote{It includes for instance the potential induced by the hidden YM instantons, which can be easily made to be negligible compared to $V(\phi)$ in Eq. (\ref{lag}). This point will be discussed in section \ref{sec:hYM}. }.
In the following we will parameterize $\gamma_\phi$ as
\dis{
\gamma_\phi \equiv \xi \frac{T^3}{f_h^2} \label{gamma}
}
with a dimensionless coefficient $\xi$ given by
\dis{
\xi \sim (N \alpha_h)^5 \left(\frac{T_h}{T} \right)^3 \label{xi},
}
where $T$ is the temperature of the dominant radiation component of the universe.
In this work, we will assume that the parameter $\xi$ is nearly a constant, which would be justified if the energy density of the axion field is negligible compared to that of the hidden thermal sector.
Later we will show that this is a valid assumption
in reasonable parameter space. 
We also ignore the running of $\alpha_h$ which would result in logarithmic-dependence of $\alpha_h$ on the temperature. As we will discuss later, this is justified as far as we are concerned with the thermal friction effect on the axion field evolution, since the thermal friction affects the axion density mostly near the time when the axion field begins to oscillate.

As for the axion potential, we will consider a cosine potential whose temperature-dependence can be parameterized as
\dis{
V(\phi) = \frac{m_0^2 f^2}{1+ (T/\Lambda_\phi)^b} \left(1 - \cos \frac{\phi}{f} \right), \label{pot}
} 
where $m_0$ is the axion mass at zero-temperature, $\Lambda_\phi$ is a dynamical scale around which this axion potential is generated by certain dynamics, and $b$ is a constant. Around its minimum, the potential can be approximated by the quadratic term with temperature-dependent axion mass
\dis{
m_\phi^2(T) \equiv \frac{m_0^2}{1+(T/\Lambda_\phi)^b}. \label{mphiT}
}
 
 In the early universe the thermal friction $\gamma_\phi$ can be larger than the Hubble rate $H$,
 because  $\gamma_\phi \propto a^{-3}$, while $H \propto a^{-3(1+w)/2}$,  where $a$ is the scale factor of the expanding universe, and $w$ is defined by the equation of state of the universe $p = w \rho$ with the total energy density $\rho$ and the total pressure $p$. Moreover, in the early universe, the thermal friction $\gamma_\phi$ can also be greater than the axion mass $m_\phi$. Using the quadratic approximation for the axion potential,
solution to Eq. (\ref{eom}) during this period (i.e. $\gamma_\phi \gg H, m_\phi$) is found to be
\dis{
\phi(t) \simeq \,&f \theta_i \exp\left( -\frac{1}{3(3+w)/2+\beta(t)} \frac{m_\phi^2(t)}{\gamma_\phi(t) H(t)} \right) \\
&\hskip -0.3cm \times \Big[ 1+ {\cal O}\left(m_\phi^2/\gamma_\phi^2, H/\gamma_\phi, \dot{\beta}/H \right) \Big]\quad
(\gamma_\phi \gg H, m_\phi), \label{phisol1}
}
where 
\dis{\phi_i\equiv f \theta_i}
 is the initial value of the axion field and 
\dis{
\beta(t) \equiv -T \frac{\partial \ln m_\phi^2}{ \partial T} = \frac{b}{1+(\Lambda_\phi/T)^b}.
}
The velocity of the axion field during this overdamping phase turns out to be
\dis{
 \dot{\phi} \simeq - \frac{m_\phi^2}{\gamma_\phi} \phi. \label{phispd1}
}
This shows that
the axion field begins to move substantially when $\gamma_\phi H$ drops below $m_\phi^2$ as the temperature decreases. 

As the universe further cools down, thermal friction eventually becomes smaller than the axion mass, and then the axion starts to oscillate around the potential minimum.
For this period (i.e.  $m_\phi \gtrsim \gamma_{\phi}, H$), we find that approximate solution to Eq. (\ref{eom}) is given by 
\dis{
\phi (t) &\simeq \phi_1 \sqrt{1+\left(\frac{\dot{\phi}_1}{m_1 \phi_1}\right)^2}  \left(\frac{m_1}{m_\phi(t)}\right)^{1/2}  \\
&\times  \exp\left(- \frac{1}{3(1-w)}\frac{\gamma_1}{H_1} \left[ 1- \left(\frac{a_1}{a}\right)^{\frac32(1-w)} \right]\right) \\
&\times  \left(\frac{a_1}{a}\right)^{3/2}  \cos \left( \int_{t_1}^{t} dt' m_\phi(t') -\tan^{-1}\frac{\dot{\phi}_1}{m_1 \phi_1}\right) \\ 
&\times \Big[1 + {\cal O}\left(\gamma_\phi^2/m_\phi^2, H^2/m_\phi^2 \right)\Big]\quad 
(m_\phi \gtrsim \gamma_{\phi}, H), \label{sol2}
}
where $t_1$ denotes the starting time of the oscillation, which is defined by \dis{m_\phi (t_1) = A \max(\gamma_\phi (t_1), 3 H(t_1))} for some constant $A \sim {\cal O}(1)$, and 
$a$ is the scale factor of the expanding universe.
The subscript $1$ means that the corresponding quantity is evaluated at $t=t_1$, for instance
\dis{m_1 \equiv m_\phi(t_1),\quad \gamma_1 \equiv \gamma_\phi(t_1),\quad \phi_1 \equiv \phi (t_1).} 

For the solutions Eq. (\ref{phisol1}) and Eq. (\ref{sol2}), we have assumed $\gamma_\phi \propto a^{-3}$. If we take into account the logarithmic temperature-dependence of $\alpha_h$, $\gamma_\phi$ also logarithmically depends on the scale factor $a$ as
\dis{
\frac{d \ln \gamma_\phi}{d \ln a} = -3 + \zeta(t),
}
where
\dis{
\zeta(t) \simeq \frac{55N}{6\pi} \alpha_h(t) \sim g_h^2(t)
}
for hidden $SU(N)$ YM gauge fields. 
It can be shown that this changes the exponential factors in
 Eq. (\ref{phisol1}) and Eq. (\ref{sol2}) as
\dis{
& \qquad \exp\left( -\frac{1}{3(3+w)/2+\beta(t)} \frac{m_\phi^2(t)}{\gamma_\phi(t) H(t)} \right) 
\\
&\rightarrow \,\, \exp\left( -\frac{1}{3(3+w)/2+\beta(t)-\zeta(t)} \frac{m_\phi^2(t)}{\gamma_\phi(t) H(t)} \right) \label{change1}
} and
\dis{
& \qquad \exp\left(- \frac{1}{3(1-w)}\frac{\gamma_1}{H_1} \left[ 1- \left(\frac{a_1}{a}\right)^{\frac32(1-w)} \right]\right)\\
& \rightarrow \,\, \exp\left(- \frac{1}{3(1-w)-2\zeta}\frac{\gamma_1}{H_1} \left[ 1- \left(\frac{a_1}{a}\right)^{\frac32(1-w)-\zeta} \right]\right). \label{change2}
}
This shows that the effects of the logarithmic running of $\alpha_h$ on the axion field evolution 
can be ignored if $g_h^2(t) \ll 1$, which can be assumed in our case since the hidden YM confinement occurs at a later time than the period when the thermal friction is active. Even when $g_h^2 = {\cal O}(1)$ over the relevant period, 
the effects due to the changes (\ref{change1}) and (\ref{change2}) can be absorbed into a factor of few  change of $f_h$ (note that $\gamma_\phi \propto f_h^{-2}$), and it does not make a qualitative change of the result. We will thus ignore the logarithmic temperature dependence of $\alpha_h$ in the following discussions.

The solutions Eq. (\ref{phisol1}) and Eq. (\ref{sol2}) also tell us that the thermal friction effect on the axion field amplitude is mostly determined around $t=t_1$ within a few e-folds. Therefore, only the value of $\xi \propto \alpha_h^5$ near $t=t_1$ will be relevant for our discussion, and the gauge coupling running effect on $\xi$ can be ignored.

The solution Eq. (\ref{sol2}) is valid when $t > t_1$. 
For the thermal friction to make a substantial difference for axion cosmology compared to the conventional scenario with $\gamma_\phi \approx 0$,
the Hubble friction has to be smaller than the thermal friction at $t=t_1$:
\dis{
\gamma_1 > 3H_1, \label{strong_friction}}
which will be assumed in the following discussions. Otherwise, $\gamma_\phi H > m_\phi^2$ before the thermal friction falls below the Hubble expansion rate, so that the axion field nearly doesn't move (see Eq. (\ref{phisol1}), and afterward the thermal friction becomes negligible compared with the Hubble rate. Thus if $\gamma_1 < 3H_1$, the axion dynamics is essentially the same as the conventional scenario.  

\begin{figure}[t!]
 \centering
 \includegraphics[scale=0.45]{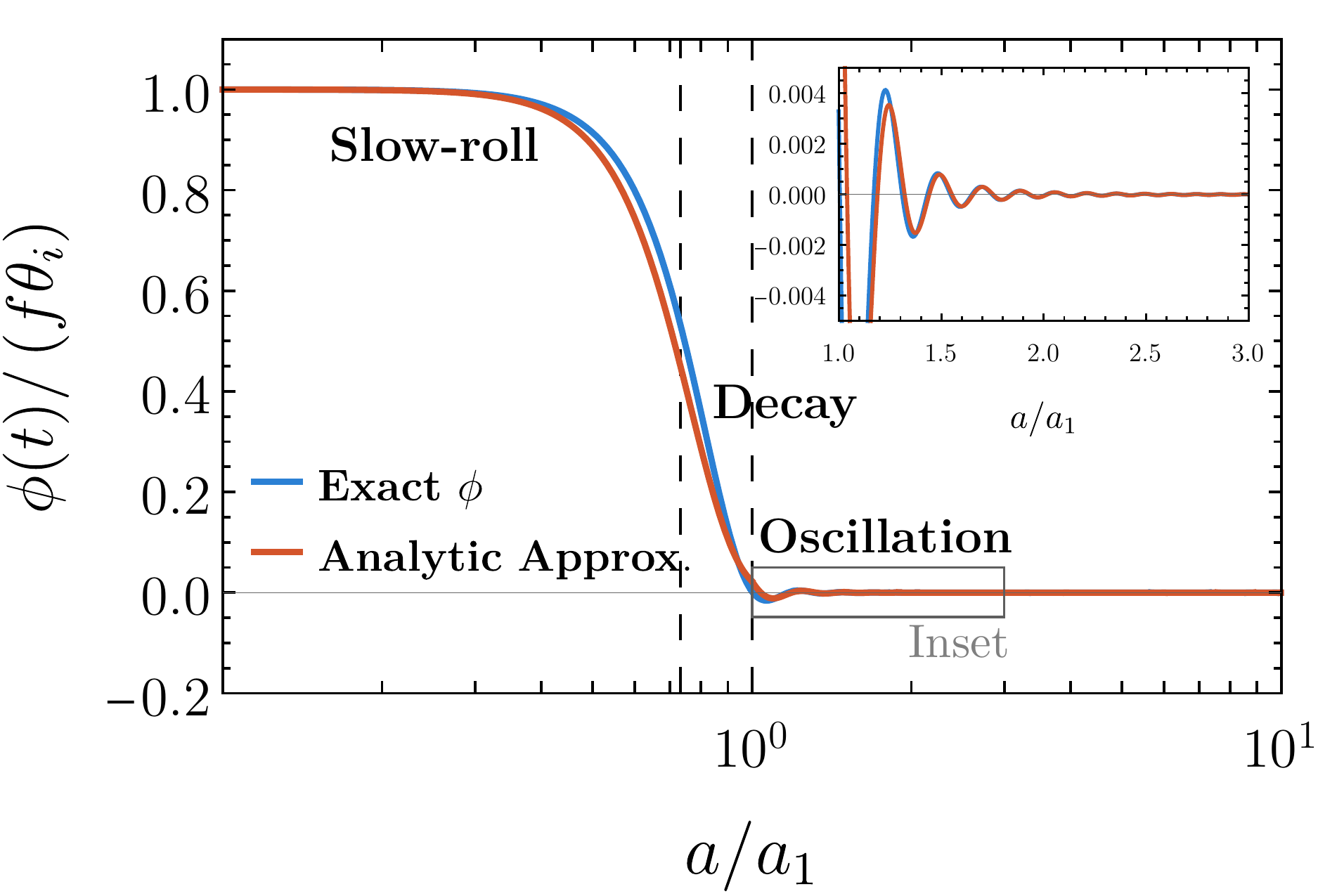}
 \caption{Cosmological axion field evolution in the presence of sizable thermal friction. Blue is the numerical solution to the axion equation of motion, and Red is the approximate analytic solution given in Eq.~(\ref{phisol1}) and Eq.~(\ref{sol2}). Here for the analytic solution, we take $A\simeq 1.0$ to get the best fit line. $a$ represents the scale factor, and $a_1$ is its value at the time when axion field begins to oscillate. At the beginning of the oscillation, $\gamma_1/H_1\simeq19$  which determines the suppression of the axion field due to thermal friction afterward (see Eq.~\ref{sol2}). Before the oscillation, the axion field already moves substantially in the Hubble time in contrast to the conventional scenario of negligible thermal friction. For the parameter set in this figure, axion abundance coincides with the observed dark matter abundance: $f_h \simeq 2.7\times 10^{6}\,{\rm GeV}$, $\xi = 10^{-7}$, $f=10^{16}\,{\rm GeV}$, $m_\phi=0.1\,{\rm eV}$, $g_{*}(T_1)=106.75$, $\theta_i=1$ and $T_1\simeq 2\,{\rm TeV}$.}
 \label{fig:osc}
\end{figure}

Let us now discuss how to connect the solution Eq. (\ref{phisol1}) for the regime $m_\phi \gtrsim \gamma_\phi$ to the solution Eq. (\ref{sol2}) for the opposite regime $m_\phi \ll \gamma_\phi$. For the case when Eq. (\ref{strong_friction}) is satisfied, which is the case of our concern, the two solutions can be connected at $t=t_1$ defined by
\dis{
m_1 \simeq A \gamma_1 \label{oscc}
}
for a constant $A \sim {\cal O}(1)$ whose precise value can be determined numerically.
The field values $\phi_1$ and $\dot{\phi}_1$ in Eq. (\ref{sol2}) are determined from the previous overdamping phase with Eq. (\ref{phisol1}) and Eq. (\ref{phispd1}):
\bea
\phi_1 &\simeq& f \theta_i \exp\left(-\frac{A^2}{3(3+w_1)/2+\beta_1} \frac{\gamma_1}{H_1}\right), \label{phiosc}  \\ 
\dot{\phi}_1 &\simeq& -A m_1 \phi_1. \label{phidotosc}
\eea
If the universe is radiation-dominated at $t=t_1$, i.e. $w_1 = 1/3$, the exponent in Eq. (\ref{phiosc}) can be determined as
\dis{
\frac{\gamma_1}{H_1} = \frac{3}{\pi} \sqrt{\frac{10}{g_*(T_1)}}\xi  \frac{M_P}{f_h^2} T_1  \label{m1t1} 
}
with the temperature $T_1$ at the onset of axion oscillation, which is given by
\dis{
T_1 \simeq \left(\frac{m_0^2 f_h^4 \Lambda_\phi^{\beta_1}}{A^2  \xi^2} \right)^{1/(6+\beta_1)} \label{oscT}
}
from the condition (\ref{oscc}).
 In Fig. \ref{fig:osc}, we depict the cosmological axion field evolution when the thermal friction is strong enough to satisfy 
Eq. (\ref{strong_friction}), connecting the solutions Eq. (\ref{phisol1}) and Eq. (\ref{sol2}) at $t=t_1$ by the conditions Eq. (\ref{phiosc}) and Eq. (\ref{phidotosc}).
We match our analytic solution to numerical solution by varying the parameter $A$ in Eq. (\ref{oscc}) and find the best fit for $A\simeq 1.0$.

During the oscillation period, the axion energy density and pressure turn out to be
\bea
\rho_\phi &\simeq& \frac12 (1+A^2) m_1 m_\phi(t) \phi_1^2  \left(\frac{a_1}{a} \right)^{3} \nonumber \\
&\times&  \exp\left(- \frac{2}{3(1-w)}\frac{\gamma_1}{H_1} \left[ 1- \left(\frac{a_1}{a}\right)^{\frac32(1-w)} \right]\right), \\ 
p_\phi &\simeq& -\rho_\phi \cos 2\left( \int_{t_1}^{t} dt' m_\phi(t') + \tan^{-1}A \right).
\eea
For $t \gg t_1$, we then find 
\bea
 \rho_\phi &\simeq& \frac12 (1+A^2) m_1 m_\phi \phi_1^2 e^{-2\gamma_1/3(1-w)H_1} \left(\frac{a_1}{a} \right)^{3} \label{rhoph} \\
\langle p_\phi \rangle_t &\simeq& 0, 
\eea
where $\langle \cdots \rangle_t$ denotes time-average over an oscillation period. It shows that eventually the axion energy density behaves like matter, i.e. $\rho_\phi \propto a^{-3}$ once $m_\phi(t) \simeq m_0$, and therefore the axion can be identified as a cold dark matter.


Let us now examine the implications of the thermal friction on axion dark matter abundance.
The axion dark matter density is determined by Eq. (\ref{rhoph}) with Eq. (\ref{phiosc}):
\dis{
 \rho_\phi \simeq m_1 m_\phi f^2 \theta_i^2 \left( \frac{a_1}{a}\right)^{3}  \exp\left(-\frac{7+\beta_1}{5+\beta_1} \frac{\gamma_1}{H_1} \right), \label{dm}
}
where we assume radiation domination at $t=t_1$ and $A\simeq 1$. 
In terms of $\Omega_\phi h^2 = \rho_\phi h^2/\rho_c $ with the critical density $\rho_c$, the abundance of the axion dark matter in the present universe is estimated as
\dis{
\Omega_\phi h^2 &\simeq 0.1\theta_i^2  \left(\frac{\xi}{10^{-7}}\right)  \left(\frac{m_0}{0.1\rm eV}\right) \left(\frac{f}{10^{16} \, \textrm{GeV}} \right)^2  \left(\frac{106.75}{g_{*s}(T_1)}\right) \\
&\times   \left(\frac{2.8\times10^{6} \,\textrm{GeV}}{f_h} \right)^2  \left(\frac{e^{-(7+\beta_1)\gamma_1/(5+\beta_1)H_1}}{10^{-11}}\right),       \label{omega}
}
where $h\equiv H_0/(100\,{\rm km}/{\rm s}/{\rm Mpc})$, and $H_0$ is the Hubble expansion rate at the present time.
Note that the above expression holds only for $\gamma_1 > 3H_1$, and therefore is not valid in the limit when the thermal friction is turned off. 
If the axion mass is independent of temperature ($b=\beta=0$), $\gamma_1/H_1$ is estimated from Eq. (\ref{m1t1}) as 
\dis{
\frac{\gamma_1}{H_1} &\simeq 18 \left( \frac{\xi}{10^{-7}}\right)^{2/3}  \left( \frac{m_\phi}{0.1\textrm{eV}}\right)^{1/3}  \\
&\times  \left(\frac{2.8\times 10^{6} \,\textrm{GeV}}{f_h}\right)^{4/3} \left( \frac{106.75}{g_*(T_1)}\right)^{1/2}. \label{m1t1num}
}
From Eq. (\ref{omega}) and Eq. (\ref{m1t1num}), we see that the axion dark matter abundance is exponentially sensitive to the axion coupling $f_h^{-1}$ for a given axion mass $m_\phi$. 
For the case $b= \beta=0$, the temperature at the onset of the axion oscillation is given by Eq. (\ref{oscT}), 
\dis{
T_1 \simeq  2 \,{\rm TeV} \,  \left(\frac{10^{-7}}{\xi} \right)^{1/3} \left(\frac{m_\phi}{0.1 \textrm{eV}} \right)^{1/3}  \left(\frac{f_h}{2.8\times 10^{6} \,\textrm{GeV}} \right)^{2/3} . \label{Toscpar}
}

A crucial feature of  the thermal friction dominated scenario defined by Eq. (\ref{strong_friction}) is  a hierarchy between the two axion scales, $f$ which would define the axion coupling to 
the dynamics generating the dominant axion potential $V(\phi)$ as
Eq. (\ref{pot})
  and $f_h$ which defines the axion coupling to the hidden YM sector as
  Eq. (\ref{lag}).
From Eq. (\ref{omega}), one can derive 
\dis{
\frac{f}{f_h} &\simeq 10^4\, \theta_i^{-1} \, \exp\left(\frac{(7+\beta_1)}{2(5+\beta_1)} \frac{\gamma_1}{H_1}\right)  \\ 
&\times \left[\left(\frac{0.1\textrm{eV}}{m_0} \right)  \left(\frac{10^{-7}}{\xi} \right) \left(\frac{g_{*s}(T_1)}{106.75} \right) \left(\frac{\Omega_\phi h^2}{0.1} \right)\right]^{1/2} \label{hierpar},
}
which tells us that  $f$ has to be much greater than $f_h$ to explain the observed dark matter abundance by the axion when thermal friction effect is important as  $\gamma_1 > 3H_1$, unless the axion is heavier than MeV. 
Such a large hierarchy may need an explanation, e.g. the clockwork mechanism \cite{Choi:2014rja, Choi:2015fiu, Kaplan:2015fuy}.

It may be instructive to compare the axion dark matter density in the presence of the thermal friction with that in the conventional scenario in which the Hubble friction dominates over other sources of friction. 
 In the conventional scenario, the onset time of the oscillation $t_1^{\rm conv}$ is determined by the condition
\dis{
3H(t_1^{\rm conv}) \simeq m_1^{\rm conv}
}
with $m_1^{\rm conv} \equiv m_\phi(t_1^{\rm conv})$, which implies
\dis{
t_1^{\rm conv} = \frac{3}{2} (m_1^{\rm conv})^{-1}
}
in the radiation-dominated era with the resulting axion dark matter density
\dis{
\rho_{\phi}^{\rm conv} = \frac12 m_1^{\rm conv} m_\phi(t) f^2 \theta_i^2 \left(\frac{t_1^{\rm conv}}{t} \right)^{3/2}.
}
Comparing this with Eq. (\ref{dm}), one can find
\dis{
  \rho_\phi &\simeq 2 \rho_\phi^{\rm conv}  \exp\left(-2\frac{(7+\beta_1)}{(5+\beta_1)} m_1 t_1\right) \\ 
  &\qquad \times \left(\frac{m_1}{m_1^{\rm conv}}\right) \left(\frac23 m_1^{\rm conv} t_1\right)^{3/2}  
\ll \rho_\phi^{\rm conv},
}
where $m_1 > m_1^{\rm conv}$ and $m_1^{\rm conv} t_1 > 3/2$ since the axion oscillation is delayed due to the thermal friction.
It shows that in our scenario $\rho_\phi$ is always smaller than $\rho_\phi^{\rm conv}$ for the same $m_\phi$ and $f \theta_i$, and exponentially suppressed in the limit $m_1 t_1 \gg 1$,
although the thermal friction delays the axion oscillation.


Finally let us discuss the consistency of the assumption that the parameter $\xi$ in Eq. (\ref{xi}) is nearly a constant. 
The hidden thermal bath actually can get a significant energy density from the axion sector. The hidden thermal bath energy density $\rho_h$ evolves according to  \cite{Berghaus:2020ekh, Berghaus:2019cls}
\dis{
\dot{\rho}_h + 4H \rho_h = \gamma_\phi \dot{\phi}^2.
}
A sufficient condition for the axion feedback to the hidden bath is negligible would be
\dis{
\gamma_\phi \dot{\phi}^2 \ll 4 H \rho_h, \label{fbcon}
}
where
\dis{
\gamma_\phi \dot{\phi}^2 \sim
\begin{cases} (m_\phi^4/\gamma_\phi) \phi^2 ~& \textrm{for} ~ \gamma_\phi > m_\phi \\
\gamma_\phi m_\phi^2 \phi^2  ~&\textrm{for}~ \gamma_\phi \lesssim m_\phi
\end{cases} 
}
from Eq. (\ref{phispd1}) and Eq. (\ref{sol2}). The solutions Eq. (\ref{phisol1}) and Eq. (\ref{sol2}) also tell us that
$\phi$ is exponentially decreasing within a single  Hubble time once $m_\phi^2/\gamma_\phi \gtrsim H$ until $t \sim t_1$, while for $t \gg t_1$, $\gamma_\phi \dot{\phi}^2 \propto 1/a^{6-\beta(t)/2}$ which implies $(m_0/m_\phi) \gamma_\phi \dot{\phi}^2 \propto 1/a^{6}$.  On the other hand, $H \rho_h \propto 1/a^{3(1+w)/2+4}$ with $3(1+w)/2+4 \leq 6$. 
Therefore, if $(m_0/m_\phi) \gamma_\phi \dot{\phi}^2 ~(\geq \gamma_\phi \dot{\phi}^2) $ is smaller than $4H \rho_h$ at the time when the axion energy density begins to decrease with $m_\phi^2/\gamma_\phi \gtrsim  H$, the condition (\ref{fbcon}) is always satisfied. 
So conservatively we may require
\dis{
\frac{m_0 m_\phi^3}{\gamma_\phi} \phi^2 \ll 4 H \rho_h
}
at $t=t_*$ defined by the condition \dis{\frac{m_\phi^2(t_*)}{\gamma_\phi(t_*)} = \left(\frac{3(3+w)}{2} +\beta\right) H(t_*)} from Eq. (\ref{phisol1}).
This requirement corresponds to
\dis{
m_0 m_* f^2 \theta_i^2 \ll r_h^4 T_*^4 \label{req}
}
where $m_* \equiv m_\phi(t_*)$, $r_h \equiv T_h/T$ which is assumed to be a constant, and we ignore coefficients of ${\cal O}(1)$.
Assuming radiation domination at $t=t_*$, the condition (\ref{req}) turns out to be equivalent to
\dis{
\frac{f}{f_h} &\ll  3\times10^{12} \theta_i^{-5/4} r_h^{5/2}  \left(\frac{m_*}{m_0}\right)^{3/8}  \\
&\times \left(\frac{10^{-7}}{\xi} \right)^{1/2} \left(\frac{1\, \textrm{TeV}}{\sqrt{m_0 f}} \right)^{1/2}. \label{req3}
} 
In section \ref{sec:hYM}, we will see that $r_h^2 \lesssim 0.1$ is required from the constraints on dark radiations.
For instance, if the axion mass is temperature-independent, that is $m_*=m_0$, one can find that the condition (\ref{req3}) is easily satisfied in the axion dark matter parameter space of Eq. (\ref{omega}) if the height of axion potential $\sqrt{m_0 f}$ is below TeV. 
As we will examine in the next section,  the QCD axion with  a temperature-dependent mass 
 also turns out to easily satisfy the condition (\ref{req3}).

\section{Implications to the QCD axion dark matter} \label{sec:qcdaxion}

For a QCD axion, in terms of Eq. (\ref{mphiT}), the temperature-dependent axion mass is given by
\dis{
\Lambda_\phi \simeq 0.15 \,\textrm{GeV}, \quad b \simeq 8
}
with
\dis{
m_{0} \simeq \frac{f_\pi m_\pi}{f} \frac{\sqrt{m_u m_d}}{m_u + m_d} \simeq 5.7 \mu\textrm{eV} \left(\frac{10^{12} \, \textrm{GeV}}{f} \right),
}
where $f_\pi$ is the pion decay constant, $m_\pi$ the pion mass, $m_q$ the bare quark mass, and $f$ is defined by the axion-gluon coupling:
\dis{
\frac{\alpha_s}{8\pi} \frac{\phi}{f} G^{\mu \nu a} \widetilde{G}^{a}_{\mu \nu}.
}

In the conventional misalignment scenario for the axion dark matter, the dark matter abundance is given by
\dis{
(\Omega_\phi h^2)_{\rm QCD}^{\rm conv} \simeq 0.1\, \theta_i^{2} \left(\frac{f}{10^{12} \, \textrm{GeV}} \right)^{7/6},
}
showing that the QCD axion dark matter is overproduced for $f \gg 10^{12} \,\textrm{GeV}$ if $\theta_i \sim {\cal O}(1)$. 
On the other hand, in our scenario in which the initial axion field exponentially decays due to  thermal friction before the onset of the oscillation, we find
(see Eq. (\ref{omega})) for the QCD axion 
\dis{
\Omega_\phi h^2 &\simeq 0.1\, \theta_i^2 \left(\frac{f}{10^{16} \, \textrm{GeV}} \right)   \left(\frac{\xi}{10^{-7}}\right)  \left(\frac{20}{g_{*s}(T_1)}\right) 
 \\
 &\times \left(\frac{0.47\times10^5\,\textrm{GeV}}{f_h} \right)^{2}   \left(\frac{e^{-15\gamma_1/13H_1}}{10^{-7}}\right),       \label{omegaqcd}
}
where 
\dis{
\frac{\gamma_1}{H_1} &\simeq 13.5 \left(\frac{\xi}{10^{-7}}\right)^{6/7} \left(\frac{0.47\times10^5\,\textrm{GeV}}{f_h} \right)^{12/7}  \\
&\times \left(\frac{20}{g_{*s}(T_1)}\right)^{1/2} \left(\frac{\Lambda_\phi}{0.15 \,\textrm{GeV}}\right)^{4/7} \left(\frac{10^{16}\,\textrm{GeV}}{f}\right)^{1/7}, \label{m1t1qcd}
}
and the temperature at the onset of the axion oscillation is given by
\dis{
T_1 \simeq \,&0.18 \,\textrm{GeV} \left(\frac{10^{-7}}{\xi} \right)^{1/7} \left(\frac{f_h}{0.47\times10^5\,\textrm{GeV}} \right)^{2/7} \\
& \times  \left(\frac{\Lambda_\phi}{0.15 \, \textrm{GeV}} \right)^{4/7}  \left(\frac{10^{16} \, \textrm{GeV}}{f} \right)^{1/7}.
}
Therefore, with the aid of thermal friction, in our scenario  the QCD axion can explain the correct dark matter abundance even for $f \gg 10^{12}  \,\textrm{GeV}$ with $\theta_i \sim {\cal O}(1)$. 
The dark matter abundance is actually very sensitive to $f_h$ due to its exponential dependence on $f_h$ as can be seen in Eq. (\ref{m1t1qcd}) and Eq. (\ref{omegaqcd}).  
For instance, if we take $f_h < 0.3\times 10^5\, \textrm{GeV}$ with 
$f=10^{16}$ GeV and $\xi=10^{-7}$, the corresponding QCD axion mass density becomes negligible as $\Omega_\phi h^2 < 6\times 10^{-9}$. On the other hand, if we take $f_h > 10^5 \, \textrm{GeV}$ instead, the thermal friction effect becomes negligible with $\gamma_1 < 3H_1$, so it recovers the conventional scenario.

The QCD axion dark matter with a large decay constant $f \gg 10^{12}$ GeV can be shown to satisfy the conservative consistency condition (\ref{req3}). 
For a QCD axion, $\sqrt{m_0 f} \sim 0.1$ GeV and 
\dis{
\frac{m_*}{m_0} \simeq \left(\frac{\Lambda_\phi}{T_*} \right)^4
}
with
\dis{
T_* &\sim 0.2 \,\textrm{GeV} \left(\frac{\Lambda_\phi}{0.15\, \textrm{GeV}} \right)^{8/13} 
\left(\frac{m_0}{0.6\, \textrm{neV}} \right)^{2/13}
\\
&\times    \left(\frac{f_h}{10^5\, \textrm{GeV}} \right)^{2/13}  \left(\frac{10^{-7}}{\xi} \right)^{1/13} 
\left(\frac{20}{g_*(T_*)}\right)^{1/26}
}
from the condition $m_\phi^2(T_*)/\gamma_\phi(T_*) \simeq 13 H(T_*)$. Therefore, the consistency condition (\ref{req3}) becomes
\dis{
\frac{f}{f_h} &\ll  2\times10^{14}\, \theta_i^{-5/4} r_h^{5/2}  \left(\frac{10^{-7}}{\xi} \right)^{1/2}
\left(\frac{0.2\,{\rm GeV}}{T_*}\right)^{3/2}
.
} 
From Eq.~(\ref{omegaqcd}) and Eq.~(\ref{m1t1qcd}), one can find that this condition is fulfilled as long as $f \lesssim 10^{18}$ GeV for a QCD axion
to explain the full dark matter abundance.


\section{Cosmological implications of hidden Yang-Mills sector} \label{sec:hYM}

A pure hidden Yang-Mills (YM) is perhaps the most straightforward and motivated example for the origin of thermal friction. In this section we discuss constraints and possible cosmological signatures from the hidden $SU(N)$ YM sector assumed in our scenario. 
The hidden YM gives rise to an axion potential through the YM instanton dynamics as well as the thermal friction through sphaleron processes.
We require that the axion potential generated by the hidden YM instantons  is small enough so that the axion dynamics discussed so far is not affected.
From the above discussion which is briefly summarized in Eq. (\ref{dm}), one can see that the thermal friction affects the axion dynamics mostly around the onset of axion oscillation with $m_1 \sim \gamma_1$. Therefore a necessary condition is that the contribution from the hidden YM to the axion mass is negligible from the time of axion oscillation. As a conservative condition, we simply require the zero-temperature axion mass from the hidden YM is smaller than the axion mass at the onset of the axion oscillation:
\dis{
\frac{\Lambda_h^2}{f_h} \ll m_1, \label{YMcd}
}
where $\Lambda_h$ is the confinement scale of the hidden YM.

For the QCD axion, on the other hand, a possible shift of the axion field vacuum value due to the
additional axion potential induced by the hidden YM dynamics has to be small enough to keep the axion solution to the strong CP problem. This requires that  
\dis{
\delta \theta \sim \frac{\Lambda_h^4/f_h}{m_\pi^2 f_\pi^2/f}\lesssim 10^{-10}  \label{strongcp}
}
where $\delta \theta=\langle \phi\rangle/f$ is the displacement of the 
QCD axion potential minimum from the CP conserving point, which has to be smaller than about $10^{-10}$ by the non-observation of the electric dipole moment of the neutron \cite{Abel:2020pzs, DiLuzio:2020wdo}.  

For an $SU(N)$ pure YM hidden gauge sector, the confinement scale is 
\dis{
\Lambda_h \simeq \mu \exp\left(- \frac{6\pi}{11N \alpha_h(\mu)}\right)
}
for a renormalization scale $\mu > \Lambda_h$. 
Because the value of $\xi \sim (N\alpha_h)^5 (T_h/T)^3$ around the onset of axion oscillation is most relevant, we may
take \dis{\mu = T_h (t_1)\equiv r_h T_1,} 
where $r_h= T_h/T$ is a constant. The condition (\ref{YMcd}) is then translated into a constraint on $\alpha_h(t_1)$:
\dis{
N \alpha_h(t_1) \lesssim \frac{\pi}{\ln(r_h^2 M_P/f_h)}\,, \label{ah1}
}
where we have assumed $\gamma_1/H_1 \lesssim {\cal O}(10) \ll r_h^2 M_P/f_h$. 
On the other hand, the condition (\ref{strongcp}) for the QCD axion becomes
\dis{
N \alpha_h(t_1) \lesssim \frac{2\pi}{\ln(r_h^4 10^{10} f/f_h)}\,. \label{ah1qcd}
}
In the interesting parameter space for the QCD axion dark matter in Eq. (\ref{omegaqcd}), the condition (\ref{ah1qcd}) is weaker than the condition (\ref{ah1}). 
If we take $r_h =0.3$, the condition  (\ref{ah1}) implies that
\dis{\xi(t_1) \sim (N\alpha_h(t_1))^5 r_h^3 \lesssim 10^{-7}} 
for $f_h > 1$ TeV.

The ratio $r_h = T_h/T$ is actually constrained by the dark radiation component made of the hidden gluons and thermalized axions. 
Currently the most stringent constraint comes from the Big Bang Nucleosynthesis (BBN) \cite{Yeh:2020mgl}:
\dis{
\Delta N_{\rm eff} = \frac{8}{7} \left(\frac{11}{4} \right)^{4/3} r_h^4 (N^2-1/2) < 0.124
}
For $SU(2)$ hidden YM, this requires $r_h \lesssim 0.3$. The constraint from the Cosmic Microwave Background (CMB) measurement is currently weaker with $\Delta N_{\rm eff} < 0.29$ \cite{Planck:2018vyg}. However, the future CMB-S4 is expected to reach $\Delta N_{\rm eff} \simeq$ 0.02-0.03 \cite{CMB-S4:2016ple, Abazajian:2019eic}, and therefore the hidden YM sector assumed in our scenario may give rise to observable dark radiation signatures in future experiments. 
Alternatively, the hidden YM sector may confine before the matter-radiation equality. Then it gives rise to self-heating hidden glueball dark matter component along with the axion dark matter \cite{Soni:2016gzf}, however this interesting possibility is beyond the scope of this paper. 

\section{Density perturbation of axion dark matter} \label{sec:denp}

The thermal friction gives rise to remarkable consequences for density perturbation of axion dark matter compared with the conventional axion cosmology.
As we have discussed so far, the axion dark matter density is exponentially reduced by the thermal friction. On the other hand, we will discuss below that the density contrast
of axion dark matter perturbation is maintained except extremely small scales. Consequently, large scale isocurvature bound on axion dark matter is relaxed in pre-inflationary Peccei-Quinn (PQ) symmetry breaking scenario. Also in the post-inflationary PQ breaking scenario, axion miniclusters can be formed at different axion mass and coupling scales compared to the conventional scenario.

In this section we will denote the spatially homogeneous part of the axion field by $\bar{\phi}(t)$,
\begin{align}
\phi(t, \mathbf{x}) = \bar{\phi}(t) + \delta \phi(t, \mathbf{x}).
\end{align}
At linear order the comoving $k$-mode of the perturbation $\delta \phi$ obeys the equation of motion
\dis{
\delta\ddot{\phi}_{\mathbf{k}} + \left(3H + \gamma_\phi\right)\delta \dot{\phi}_{\mathbf{k}}+\left(\frac{k^2}{a^2} + m_\phi^2 \right)\delta \phi_{\mathbf{k}} = \sum_i S_i \label{delphieq}
}
where $k\equiv |\mathbf{k}|$ for comoving wavenumber vector $\mathbf{k}$, and $S_i$'s are external source terms.
Note that we do not specify a gauge choice in Eq.~\eqref{delphieq}.
In the subsequent discussions, we will specify the source terms after fixing the gauge.
For the moment, let us ignore the source terms.
As we have discussed in section \ref{sec:dynamics}, the background field $\bar{\phi}$ begins to decay into the hidden thermal bath after $t=t_*$ defined by 
\dis{
m_\phi (t_*) \simeq \sqrt{\gamma_\phi(t_*) H(t_*)}, \label{tstar}
}
see Eq. (\ref{phisol1}). Hereafter, we will denote quantities evaluated at $t=t_*$ with the asterisk, e.g. $m_* \equiv m_\phi(t_*), a_*\equiv a(t_*)$.
On the other hand, for the perturbation $\delta \phi_\mathbf{k}$, 
high $k$-modes with $k/a_* \gg m_*$ begins to decay earlier than the background field 
because of the effective mass $m_{\rm eff}^2 (k) \equiv k^2/a^2 + m_\phi^2 \gg m_\phi^2$. Therefore $\delta \phi_k/ \bar{\phi}$ becomes negligible for the modes with $k/a_* \gg m_*$ for $t>t_*$. On the contrary, for low $k$-modes with $k/a_* \lesssim m_*$, $m_{\rm eff}^2(k) \simeq m_\phi^2$ for $t > t_*$, and thus $\delta \phi_k$ decays almost in the same way as the background field value $\bar{\phi}$ so that $\delta \phi_k / \bar{\phi}$ remains constant. 
The resultant density contrast $\delta \rho_\phi/\rho_\phi \sim \delta \phi/\bar{\phi}$ will be maintained until  gravitational collapse of the axion over-density.
In summary,
\dis{
\frac{\delta \phi_\mathbf{k}}{\bar{\phi}} \simeq 
\begin{cases}
0, & k/a_* \gg m_* \\
\left(\frac{\delta \phi_\mathbf{k}}{\bar{\phi}}\right)_i, & k/a_* \lesssim m_*
\end{cases} \label{cont}
}
for $t_* < t < t_c$ with $t_c$ critical time for gravitational collapse of the axion density, and $(\delta \phi_\mathbf{k}/\bar{\phi})_i$ is the initial value for the ratio before $t=t_*$.

In the present universe the corresponding physical length scale for the comoving wavenumber $k = a_* m_*$ is
\dis{
\frac{z_*}{m_*} \simeq 1\,\textrm{pc} \left(\frac{T_*}{0.1 \, \textrm{GeV}} \right) \left(\frac{10^{-11} \textrm{eV}}{m_*} \right),
} 
where $z_*+1 \equiv a_0/a_*$ with the current scale factor $a_0$. The reference values for $T_*$ and $m_*$ are typical values for the QCD axion. 
Thus in case of the QCD axion, for example, the density contrast $\delta \rho_\phi/\rho_\phi \sim \delta \phi/\bar{\phi}$ for large scales greater than 1 pc in the current universe is not diluted by the thermal friction, while the average density $\rho_\phi$ can be exponentially diluted. 
On the other hand, axion density contrast at smaller scales may be significantly affected by the thermal friction. For this case, the source terms in Eq. (\ref{delphieq}) can be important. 
In the following subsections, we will discuss
these points in detail for different PQ symmetry breaking scenarios.

\subsection{Pre-inflationary scenario}

In the pre-inflationary PQ breaking scenario, the initial axion field perturbation is determined by the inflationary Hubble scale
\dis{
\delta \phi_\mathbf{k} = \frac{H_I}{2\pi}.
}
The isocurvature power spectrum generated by this perturbation is
\dis{
P_{\rm iso} = \left( \frac{\Omega_\phi}{\Omega_{\rm DM}} \right)^2 \frac{H_I^2}{\pi^2 f^2 \theta_i^2}, \label{piso}
}
where $\Omega_{\rm DM}= \rho_{\rm DM}/\rho_c$ is the observed dark matter abundance.
The CMB measurement constrains this uncorrelated axion isocurvature perturbation by \cite{Planck:2018jri}
\dis{
\frac{P_{\rm iso}}{P_s} < 0.038,
}
where $P_s$ is the scalar perturbation measured as
\dis{
    P_s (k) &\simeq 2.1\times 10^{-9} \left(\frac{k}{0.05\,{\rm Mpc}^{-1}}\right)^{n_s-1} \label{Ps}
}
with the spectral index $n_s \simeq 0.96$~\cite{Planck:2018vyg}. From Eq. (\ref{omega}), 
\dis{
f \theta_i &\simeq 10^{11} \, \textrm{GeV}   \left( \frac{\Omega_\phi}{\Omega_{\rm DM}} \right)^{1/2} \exp \left(\frac{7+\beta_1}{2(5+\beta_1)} \frac{\gamma_1}{H_1} \right) \\ 
&\times  \left(\frac{0.1\textrm{eV}}{m_0} \right)^{1/2}  \left(\frac{10^{-7}}{\xi} \right)^{1/2}\left( \frac{f_h}{10^7\, \textrm{GeV}}\right) \left(\frac{g_{*s}(T_1)}{106.75} \right)^{1/2} \label{fthi}
}
For a given axion mass $m_0$, therefore, the isocurvature power spectrum of the axion dark matter can be exponentially reduced by the factor $e^{(7+\beta_1)\gamma_1/(5+\beta_1) H_1}$  even for $\Omega_\phi = \Omega_{\rm DM}$,
as one can see from Eq. (\ref{piso}) and Eq. (\ref{fthi}).
If $f$ is determined by axion mass $m_0$ like the QCD axion, the isocurvature bound can be still relaxed by a larger $\theta_i$ to explain the dark matter abundance.

For small scales with $k/a_0 > 0.1\, \textrm{Mpc}^{-1}$, the isocurvature perturbations of dark matter are not constrained by the CMB data \cite{Planck:2018jri}. 
It has been suggested that the small scale axion isocurvature perturbations can be enhanced by several mechanisms including tachyonic instability \cite{Fukunaga:2020mvq}, parametric resonance \cite{Arvanitaki:2019rax}, and resonance with the radiation fluctuation \cite{Sikivie:2021trt, Kitajima:2021inh} when the Hubble damping is the dominant source of friction for axion dynamics. 
Here we examine how the thermal friction would change the picture and affect the growth of small scale perturbations. 
For the small scale perturbations, the source terms in Eq. (\ref{delphieq}) can be important, and they may be classified according to the convention of \cite{Kitajima:2021inh}:
\bea
S_1 &=& -\left(\dot{\Psi}_{\mathbf{k}}+3\dot{\Phi}_{\mathbf{k}}\right)\dot{\bar{\phi}}, \label{s1}\\
S_2 &=& 2\Psi_{\mathbf{k}} m_\phi^2(T) f \sin\frac{\bar{\phi}}{f}, \label{s2}\\
S_3 &=& -(\partial_T m_\phi^2)\, \delta T_{\mathbf{k}} f \sin \frac{\bar{\phi}}{f}, \label{s3}\\
S_4 &=& \left(\Psi_{\mathbf{k}} \gamma_\phi - (\partial_{T_h} \gamma_\phi) (\delta T_h)_{\mathbf{k}} \right)\dot{\bar{\phi}}, \label{s4}
\eea
where $\delta T$ and $\delta T_h$ are the temperature fluctuations of the visible radiation and the hidden radiation, respectively, and the metric perturbations $\Psi$ and $\Phi$ are defined by
\dis{
ds^2  =
 -a^2(\tau) \left[\left(1- 2\Psi(\tau, \mathbf{x})\right)d\tau^2
 -\left(1+ 2\Phi(\tau, \mathbf{x})\right)d\mathbf{x}^2
\right], \label{eq:metric}
}
in the conformal Newtonian gauge~\cite{Mukhanov:1990me,Ma:1995ey}.
Here $S_1$ and $S_2$ are source terms for axion perturbation from gravitational potentials, $S_3$ from the radiation fluctuation via temperature-dependent axion mass, and $S_4$ is a new source term originated from the thermal friction.\footnote{In principle, there can be additional thermal terms beyond the fricition term, caused by the inhomogeneity of the axion field. However, that terms turn out to be negligible in small momentum expansion of the axion field (c.f. \cite{McLerran:1990de}). }

Assuming that the anisotropic stress of the energy-momentum tensor vanishes, we have 
\dis{
\Psi = \Phi.
}
The analytic solutions for the gravitational potential $\Phi$ and the temperature fluctuation $\delta T$ in the radiation-dominated era are given by~\cite{Dodelson:2003ft}
\begin{align}
\Phi_{\mathbf{k}}&= 3(\Phi_{\mathbf{k}})_i \frac{\sin y_k-y_k \cos y_k}{y_k^3},
\label{eq:Phik}
\\
\frac{\delta T_{\mathbf{k}}}{T}
&=
\left(
\frac{1}{2}+\frac{1}{2} y_k^2
\right)
\Phi_{\mathbf{k}}
+
\frac{1}{2H} \dot{\Phi}_{\mathbf{k}}
\,,
\label{eq:deltaTk}
\end{align}
where $(\Phi_{\mathbf{k}})_i$ is the primordial Newtonian potential, and
$y_k \equiv k/(\sqrt{3} a H)$.
For the analytic solutions, we have assumed that $r_h^4=T_h^4/T^4 \ll 1$. By this assumption, it can be also shown that
\dis{
 \frac{(\delta T_h)_{\mathbf{k}}}{T_h} \simeq \frac{\delta T_{\mathbf{k}}}{T},
}
which will be assumed in the following analysis.

The axion isocurvature perturbation is defined by
\dis{
    S_{\phi}&\equiv -3H \left(\frac{\delta\rho_\phi}{\dot{\rho}_\phi}-\frac{\delta\rho_r}{\dot{\rho}_r}\right) \simeq \frac{\delta\hat{\rho}_{\phi}}{\rho_\phi} \quad (t \gg t_1) \label{sphi}
 }
where $\rho_r$ is the radiation density, and
\dis{
\delta \hat{\rho}_\phi \equiv \delta \rho_\phi - \frac{\dot{\rho}_\phi}{\dot{\rho}_r} \delta \rho_r 
}
which is a gauge-invariant quantity. For $t \gg t_1$, we have used $\dot{\rho}_\phi + 3H\rho_\phi \simeq 0$ in Eq. (\ref{sphi}). 
In the conformal Newtonian gauge, $\delta \hat{\rho}_\phi$ is given by
\begin{align}
    \delta\hat{\rho}_{\phi}
    &=
    \dot{\bar{\phi}}\,\delta\dot{\hat{\phi}} + \partial_{\bar{\phi}} V \delta\hat{\phi} + \hat{\Psi} \dot{\bar{\phi}}^2
    ,
    \label{eq:deltarhophirad}
    \\
    &=
    \dot{\bar{\phi}}\delta\dot{\phi}+m_\phi^2 f\sin\frac{\bar{\phi}}{f}\, \delta\phi-\left(\left(3+\frac{\gamma_\phi}{H}\right)\frac{\delta T}{T}-\Psi\right)\dot{\bar{\phi}}^2
    ,
    \label{eq:deltarhophiradsimp}
\end{align}
where
\begin{align}
    \delta\hat{\phi}
    &\equiv
    \delta\phi - \frac{\dot{\bar{\phi}}}{\dot{\rho}_r}\delta\rho_r,
    \\
    \hat{\Psi}
    &\equiv
    \Psi+\partial_t \left( \frac{\delta\rho_r}{\dot{\rho}_r} \right)
    .
\end{align}
We numerically calculate the power spectrum for $\delta_\phi \equiv \delta \hat{\rho}_\phi/\rho_\phi$, which is equivalent to the power spectrum
of the axion isocurvature perturbation $S_\phi$ for $t \gg t_1$, i.e.
\dis{
    \langle \delta_\phi (\mathbf{k}) \delta_\phi(\mathbf{k'})^\ast \rangle \simeq (2\pi)^3 \delta^{(3)} (\mathbf{k}-\mathbf{k'}) \frac{2\pi^2}{k^3} 
    P_{\rm iso} (k).
 }
 
 \begin{figure*}[th]
\centering
\includegraphics[width=0.45\textwidth]{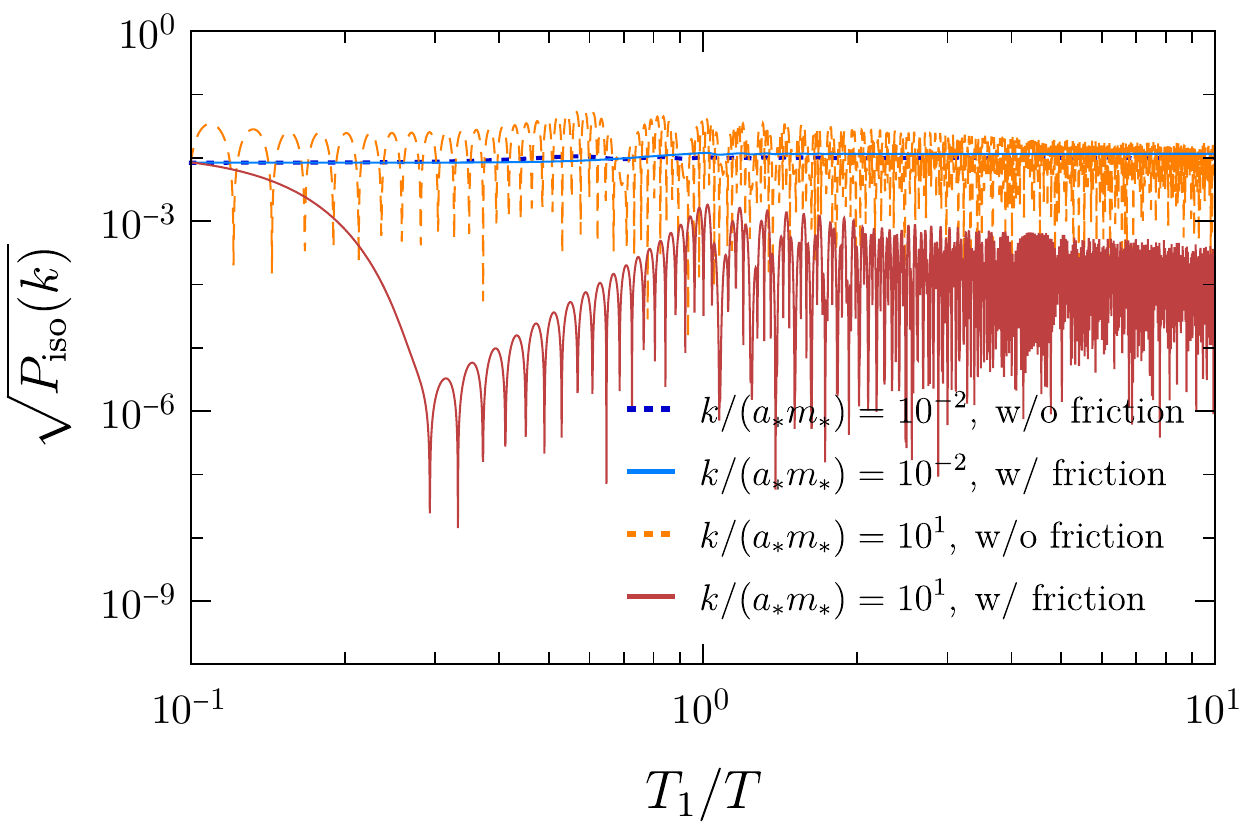} \hspace{1cm}
\includegraphics[width=0.45\textwidth]{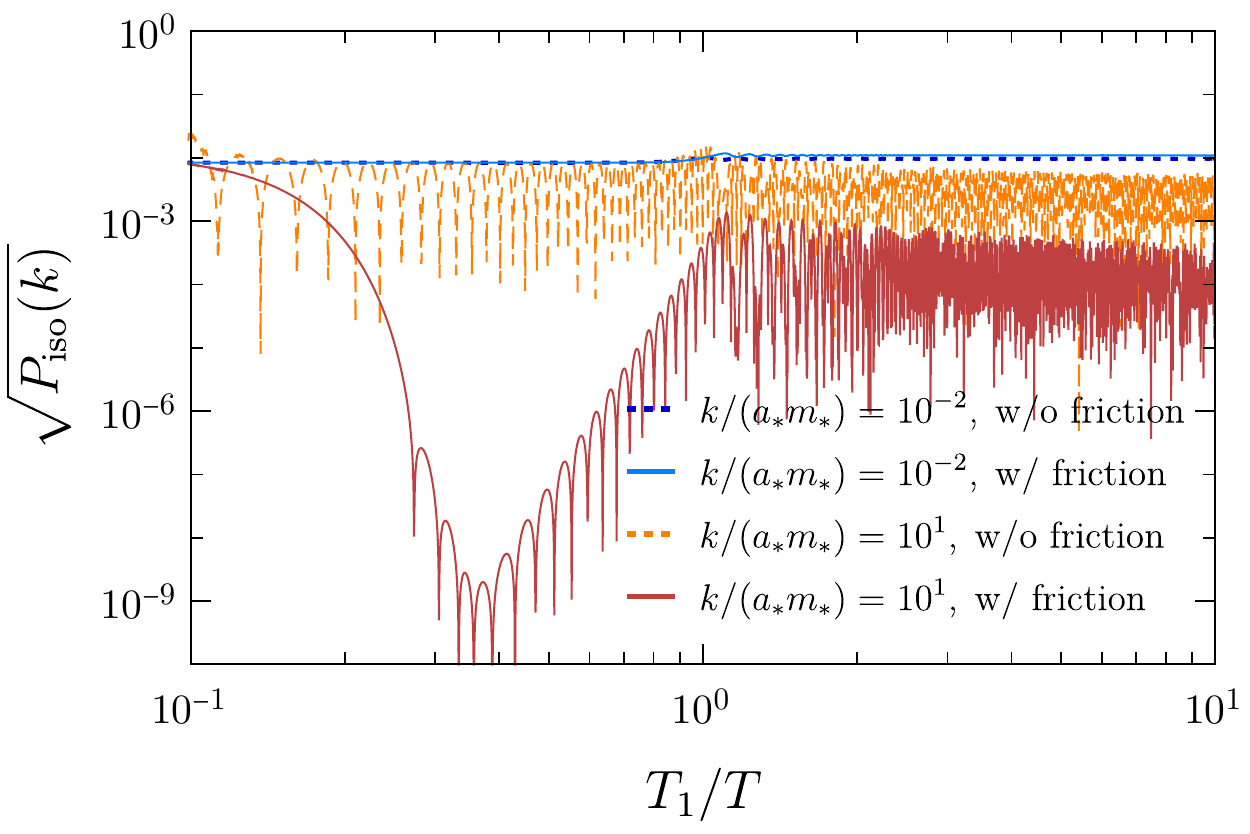} 
\caption{({\it Left}) Evolution of power spectrum of $\delta_\phi = \delta\hat{\rho}_{\phi}/\rho_\phi$ with a temperature-independent axion mass ($b=0$) for a large initial field perturbation $(\delta\hat{\phi}/\phi)_i=10^{2} \sqrt{2.1 \times 10^{-9}}$. 
The solid (dashed) lines represent the evolution with (without) thermal friction.
The low-$k$ modes, i.e., $k / (a_\ast m_\ast) \lesssim 1$, maintain their initial amplitudes since $m^2_{\rm eff}(k) \simeq m^2_\phi$ and the source terms are suppressed \cite{Kitajima:2021inh}.
The high $k$-modes with thermal friction (red-solid line), i.e., $k / (a_\ast m_\ast) \gg 1$, exhibit the initial decay state due to their large effective mass.
The decay stage is followed by a mild enhancement of $\sqrt{P_{\rm iso}}$ approaching the order of the curvature perturbation, $\delta T/ T \sim \sqrt{P_s}$.
Here the parameter set is chosen to explain the total dark matter abundance by the axion in the presence of thermal friction: $f_h\simeq 3\times 10^6 \,{\rm GeV}$, $\xi=10^{-7}$, $f=10^{16}\,{\rm GeV}$, $m_\phi=0.1\,{\rm eV}$,
$g_\ast(T_1)=106.75$, $\theta_i=1$, and $T_1\simeq 2\,{\rm TeV}$.
({\it Right}) 
Same as the left panel but for the QCD axion ($b\simeq8$). 
The qualitative behaviors are similar to those in the temperature-independent axion mass case.
Here the parameter set is chosen to explain the total dark matter abundance by the axion in the presence of thermal friction: $f_h\simeq 0.5\times 10^5 \,{\rm GeV}$, $\xi=10^{-7}$, $f=10^{16}\,{\rm GeV}$, $m_\phi=0.6\,{\rm neV}$, 
$g_\ast(T_1)=20$, $\theta_i=1$, and $T_1\simeq 0.18\,{\rm GeV}$.
}
\label{fig:Tindep}
\end{figure*}


 In Fig.~\ref{fig:Tindep}, we show our results for the power spectrum of $\delta_\phi (\mathbf{k})$.
The left panel represents the case for $T$-independent axion mass $(b=0)$, while the right panel is for the QCD axion case.  
We choose a large initial field perturbation $(\delta\hat{\phi}/\phi)_i=10^{2} \sqrt{2.1 \times 10^{-9}}$ to demonstrate the effect of the thermal friction. 
The solid (dashed) lines represent the evolution of density perturbation with (without) thermal friction.
The low-$k$ modes, i.e., $k / (a_\ast m_\ast) \lesssim 1$, maintain their initial amplitudes since $m^2_{\rm eff}(k) \simeq m^2_\phi$, and the source terms are suppressed \cite{Kitajima:2021inh}.
For the high $k$-modes (red-solid line), i.e., $k / (a_\ast m_\ast) \gg 1$, due to their large effective masses, $\delta\phi_\mathbf{k}$ oscillates prior to the oscillation of the background field $\bar{\phi}$. Consequently, the thermal friction  renders the initial decay stage of the high-$k$ modes erasing the axion isocurvature power spectrum $\sqrt{P_{\rm iso}}$. The decay stage is followed by a mild enhancement stage due to the source terms in Eq.~\eqref{delphieq}, during which $\sqrt{P_{\rm iso}}$ becomes comparable to the curvature perturbation, $\delta T/ T \sim \sqrt{P_s}$.

Previous works~\cite{Sikivie:2021trt,Kitajima:2021inh} have pointed out that the terms proportional to $\delta \phi$ in Eq.  (\ref{eq:deltarhophiradsimp}) can be resonantly enhanced by the source term $S_3$ for which the temperature dependence of the axion mass is crucial.
We find that such resonance effect is negligible for light QCD axion around neV, where sizable thermal friction is required to explain the dark matter abundance by the QCD axion.  The thermal friction increases the width of the resonance, making the resonant amplitude smaller.  Furthermore, we have examined the possibility of the parametric resonance which can happen when the initial axion field value is close to the hilltop of the axion potential~\cite{Arvanitaki:2019rax,Fukunaga:2020mvq,Kitajima:2021inh}. 
In the presence of sizable thermal friction, however, we find such resonant growth is also hindered. This is because the axion field amplitude is quickly damped down to the regime where the non-linear axion self-interaction becomes negligible.



\subsection{Post-inflationary scenario}
 
In the post-inflationary scenario, the axion field takes random values over the universe after the PQ phase transition. 
In the presence of the thermal friction, those random field fluctuations are diluted only for small length scales with $k/a \gtrsim \sqrt{\gamma_\phi H}$ 
as observed in Eq. (\ref{delphieq}) and Eq. (\ref{phisol1}), which are smaller than the horizon scale when $\gamma_\phi > H$. 
This is essentially because the thermal friction holds the field values by stronger friction than the Hubble damping.
Therefore, a global string can exist in the following length scale smaller than the horizon scale:
\dis{
d_{\rm string} \sim 1/\sqrt{\gamma_\phi H} < 1/H.
}
It means the behavior of the scaling solution \cite{Kibble:1976sj, Albrecht:1984xv, Bennett:1987vf, Gorghetto:2018myk} for strings is changed due to the thermal friction. 

The string-wall network is formed and collapses around the time when the background axion field starts to substantially move 
 after $t=t_*$ defined by Eq. (\ref{tstar}).\footnote{Here we assume the domain wall number $N_{\rm DW}=1$. Otherwise the string-wall network can be collapsed long after $t=t_*$ depending on the lifetime of the domain walls.} Thus ${\cal O}(1)$ density contrast of the axion dark matter will be formed at the scale $k_c/a_* \sim \sqrt{\gamma_* H_*} \simeq m_*$ around $t=t_*$. This ${\cal O}(1)$ density contrast is not diluted by the thermal friction according to the argument given around Eq. (\ref{cont}), while the average axion density can be exponentially diluted. If the axion is a dominant component of the total dark matter, 
 the axion clumps are expected to collapse gravitationally prior to matter-radiation equality into small dark matter halos, called axion miniclusters \cite{Hogan:1988mp, Kolb:1993zz, Vaquero:2018tib}.
The typical axion minicluster size is estimated to be
\dis{
r_{\rm mc} \lesssim \left(\frac{a_{\rm eq}}{a_*}\right) m_*^{-1} \label{mcsize}
}
where $a_{\rm eq}$ is the scale factor at the matter-radiation equality.
This means that the thermal friction makes the minicluster size smaller compared to the conventional scenario for a given axion mass, because the string-wall network collapses later due to stronger friction.

For the QCD axion the estimation (\ref{mcsize}) implies an interesting consequence. 
As we have discussed in section \ref{sec:qcdaxion}, the QCD axion scale $f$ can be much larger than the intermediate scale while explaining the dark matter abundance without a small $\theta_i$ thanks to the strong damping by the thermal friction. Thus the QCD axion can be far lighter than meV scale even in the post-inflationary scenario without overproducing axion dark matter. According to Eq. (\ref{mcsize})
 the corresponding axion minicluster size is to be much larger than what is expected in the conventional scenario. 
 
\section{Conclusions} \label{sec:conc}

Axion field dynamics in the early universe might be significantly affected by  interactions with a hidden thermal bath. 
A well motivated example is an axion coupling to hidden YM gauge fields. The sphaleron processes then give rise to a friction term for the cosmological evolution of axion field.
In this work we have examined the influence of such a thermal friction to the abundance of axion dark matter and its density perturbations.

Assuming that the energy density of the hidden thermal bath is large enough compared with the axion energy density,
the temperature $T_h$ of hidden thermal bath
is maintained to be proportional to the temperature $T$ of the universe. This gives rise to thermal friction approximately proportional to $T^3$. 
We have derived approximate analytic solutions to the equation of motion of the axion field for the case that
thermal friction remains to be greater than the Hubble expansion rate until the onset of the axion field oscillation around the potential minimum. Our solutions show that there is an exponential decay phase of the axion field before the axion begins to oscillate. Because of this decay phase, the axion dark matter abundance can be exponentially reduced by the factor $\sim e^{-\gamma_1/H_1}$,   where $\gamma_1$ and $H_1$ denote the thermal friction and the Hubble expansion rate, respectively, when the axion field begins to oscillate. This allows that
the model parameter region which would yield an overclosing relic axion mass density in
the conventional scenario without thermal friction is cosmologically viable in the presence of thermal friction. 
In particular, it opens a new parameter space for the QCD axion dark matter with  $m_\phi \ll \mu$eV without having a fine-tuned initial misalignment angle. The dark radiation formed by hidden gluons and thermalized axions may also result in potentially observable signatures in the future CMB measurements.

A crucial ingredient of this scenario is a big hierarchy among the two axion scales, $f$ characterizing the axion coupling to the dynamics generating the dominant axion potential and $f_h$ characterizing the axion coupling to  generate the thermal friction. Such a scale (or coupling) hierarchy might be naturally generated by the clockwork mechanism \cite{Choi:2014rja, Choi:2015fiu, Kaplan:2015fuy}.

Thermal friction can also make remarkable consequences for the density perturbations of axion dark matter. In the pre-inflationary PQ breaking scenario, the large-scale isocurvature perturbation generated during the early universe inflation experiences the same exponential decay as the
average axion field value does. As a consequence, for a given axion dark matter relic density, the 
isocurvature bound on the inflationary energy scale can be significantly ameliorated in our scenario
with strong thermal friction.  The thermal friction also suppresses the power spectrum of axion density perturbation at small scales below the scale determined by the axion mass at the onset of the exponential decay. It turns off various resonance effects which can be present in the absence of thermal friction. In the post-inflationary PQ breaking scenario, on the other hand, the thermal friction keeps the random axion field values in a scale smaller than the horizon size. Thus it changes the scaling behavior of the axionic strings making them to be present by $O(1)$ population in a scale significantly smaller than the horizon size. 
As a consequence, the typical axion minicluster size is to be smaller than the conventional one for a given axion mass. For the QCD axion, combined with the fact that the thermal friction allows a larger decay constant well above $10^{10}$ GeV even in the post-inflationary PQ breaking scenario, the axion minicluster size can be much greater than the expected size in the conventional axion cosmology without thermal friction effect.
 \\
 \\
 \noindent{\bf Acknowledgements}
\\
\\
This work was supported by IBS under the project code, IBS-R018-D1. Hyeonseok Seong was also supported by the Deutsche Forschungsgemeinschaft under Germany Excellence Strategy — EXC 2121 “Quantum Universe” — 390833306.  We thank Chang Sub Shin and Jai-chan Hwang for helpful discussions.
\\
\\
 \noindent{\bf Note added:} Our discussion of the axion dark matter abundance largely overlaps with Ref. \cite{Papageorgiou:2022prc}, whose first version was put forward on the arXiv a few days before ours. Ref. \cite{Papageorgiou:2022prc} also noticed a possibility that axion dark matter abundance is enhanced by thermal friction compared to the conventional scenario due to spontaneous breaking of the hidden gauge symmetry right before the decay phase, which is not addressed in this work. On the other hand, our work includes a study of the effect of thermal friction on density perturbation of axion dark matter, which is not covered in Ref. \cite{Papageorgiou:2022prc}. We note that the motivation of our work was influenced by a journal club talk at IBS-CTPU presented by A. Papageorgiou on January 27th, 2022, in which he shared preliminary results of Ref. \cite{Papageorgiou:2022prc}.

%


 \end{document}